\documentclass[%
superscriptaddress,
nofootinbib,
 amsmath,amssymb,
 aps,
 pra,
floatfix,
twocolumn
]{revtex4-1}
\usepackage[utf8]{inputenc}
\usepackage[T1]{fontenc}
\usepackage{babel}
\usepackage{graphicx}
\usepackage{amsfonts}
\usepackage{amssymb}
\usepackage{amsmath}
\usepackage{amsthm}
\usepackage{mathtools}
\usepackage{physics}
\usepackage[normalem]{ulem}
\usepackage{blindtext}
\usepackage{dsfont}

\usepackage{mathtools}
\usepackage{braket}

\usepackage{graphicx}
\usepackage[format=plain,justification=centerlast]{caption}
\usepackage[format=plain,justification=centerlast]{subcaption}
\usepackage{bm}

\usepackage[dvipsnames]{xcolor}

\usepackage{natbib}
\setcitestyle{square,numbers}
\usepackage[colorlinks]{hyperref}

\begin{document}

\author{Tomasz Linowski}
\affiliation{International Centre for Theory of Quantum Technologies, University of Gdansk, 80-308 Gda{\'n}sk, Poland}
\email[Corresponding author: ]{t.linowski95@gmail.com}

\author{Konrad Schlichtholz}
\affiliation{International Centre for Theory of Quantum Technologies, University of Gdansk, 80-308 Gda{\'n}sk, Poland}

\author{Giacomo Sorelli}
\affiliation{Laboratoire Kastler Brossel, Sorbonne Université, ENS-Université PSL, CNRS, Collège de France, 4 Place Jussieu, 75252 Paris, France}
\affiliation{Fraunhofer IOSB, Ettlingen, Fraunhofer Institute of Optronics,
System Technologies and Image Exploitation, Gutleuthausstr. 1, 76275 Ettlingen, Germany}

\author{Manuel Gessner}
\affiliation{Departamento de F\'{i}sica Te\'{o}rica and IFIC, Universidad de Valencia-CSIC, C/ Dr. Moliner 50, 46100 Burjassot (Valencia), Spain}

\author{Mattia Walschaers}
\affiliation{Laboratoire Kastler Brossel, Sorbonne Université, ENS-Université PSL, CNRS, Collège de France, 4 Place Jussieu, 75252 Paris, France}

\author{Nicolas Treps}
\affiliation{Laboratoire Kastler Brossel, Sorbonne Université, ENS-Université PSL, CNRS, Collège de France, 4 Place Jussieu, 75252 Paris, France}

\author{{\L}ukasz Rudnicki}
\affiliation{International Centre for Theory of Quantum Technologies, University of Gdansk, 80-308 Gda{\'n}sk, Poland}
\affiliation{Center for Theoretical Physics, Polish Academy of Sciences, 02-668 Warszawa, Poland}
\email[Corresponding author: ]{lukasz.rudnicki@ug.edu.pl}

\title{Application range of crosstalk-affected spatial demultiplexing for resolving separations between unbalanced sources}

\date{\today}

\begin{abstract}
Superresolution is one of the key issues at the crossroads of contemporary quantum optics and metrology. Recently, it was shown that for an idealized case of two balanced sources, spatial mode demultiplexing (SPADE) achieves resolution better than direct imaging even in the presence of measurement crosstalk [Phys. Rev. Lett. \textbf{125}, 100501 (2020)]. In this work, we consider arbitrarily unbalanced sources and provide a systematic analysis of the impact of crosstalk on the resolution obtained from SPADE. As we dissect, in this generalized scenario, SPADE's effectiveness depends non-trivially on the strength of crosstalk, relative brightness and the separation between the sources. In particular, for any source imbalance, SPADE performs worse than ideal direct imaging in the asymptotic limit of vanishing source separations. 
%\tb{In this work, we consider the more general case of unbalanced sources and provide a systematic analysis of the impact of crosstalk on the resolution obtained from SPADE depending on the strength of crosstalk, relative brightness and the separation between the sources. We find that, in contrast to the original findings for perfectly balanced sources, SPADE performs worse than ideal direct imaging in the asymptotic limit of vanishing source separations.} 
Nonetheless, for realistic values of crosstalk strength, SPADE is still the superior method for several orders of magnitude of source separations.
\end{abstract}

\maketitle

\section{Introduction}

In recent years, considerable attention was devoted to the problem of resolving asymptotically small separations between two point-like light sources. In the case of separations below the so-called Rayleigh regime, which is determined by the optical apparatus' point spread function width \cite{resolution_survey_denDekker_1997,Fourier_optics_Goodman_2005}, the efficiency of traditional measurement schemes relying on direct imaging drops significantly \cite{Rayleigh_curse_Paur_2018}.

Over the course of the last two decades, a plethora of methods was developed to overcome the limitations of direct imaging and achieve superresolution. This includes photoactivated localization microscopy (PALM) \cite{PALM_Betzig_2006,PALM_Hess_2006}, optical reconstruction microscopy (STORM) \cite{STORM_Hell_2007}, the use of superoscillations \cite{superoscillations_Smith_2016,superoscillations_Gbur_2019} and inversion of coherence along an edge (SPLICE) \cite{SPLICE_Tham_2017,SPLICE_Bonsma-Fisher_2019}, among others \cite{superresolution_techniques_Hemmer_2012,superresolution_techniques_Liang_2021}. In the particular case of estimating the distance between two incoherent light sources, such as a planet orbiting around a distant star \cite{hypothesis_testing_exoplanets_2021}, the optimal measurement is given by \emph{spatial demultiplexing (SPADE)} in Hermite-Gauss modes \cite{superresolution_Tsang_2016,superresolution_starlight_Tsang_2019}. 

Unfortunately, due to the presence of noise and technical imperfections, such as apparatus misalingment, no measurement scheme is ideally implemented in experimental setups \cite{noon_advances,metrology_noise_Kolodynski_2013,metrology_noise_Nichols_2016,noise_Oh_2021,Sorelli_moment-based_superresolution_2021,Sorelli_practical_superresolution_2021,imaging_noisy_Lupo_2020,superresolution_limits_SPADE_Len_2020,resolution_misalingment_Almeida_2021,superresolution_no_location_Grace_2020}. In the case of SPADE, there is always a small fraction of the measured mode that is not transmitted into the correct output, but to another mode instead: a phenomenon known as crosstalk. Recent efforts showed that, while crosstalk lowers SPADE's efficiency significantly, the method is still typically superior even to ideal (i.e. noiseless and continuous) direct imaging \cite{crosstalk_original_PRL}. These findings, however, were based on the assumption that the two light sources are exactly equally bright, which is often not the case, as in the aforementioned example of a planet-star system.

In this work, we assess the applicability of SPADE to resolving separations between unbalanced sources, i.e. sources of arbitrary relative brightness \cite{unbalanced_sources_Rehacek_2017}, in practical scenarios. As we find, the case of perfectly balanced sources is the only one, for which SPADE is always more efficient than direct imaging. 
Otherwise, the effectiveness of SPADE has a non-trivial dependence on the relation between measured separations (normalized with respect to the point spread function width) and crosstalk strength, which we investigate in detail.
%\tb{Otherwise, the effectiveness of SPADE depends on the relation between measured separations (normalized with respect to the point spread function width) and crosstalk strength.} 
Using analytical tools, we show that for asymptotically vanishing distances, SPADE is outperformed by direct imaging, while for distances much larger than crosstalk strength (which are still below the Rayleigh regime), SPADE remains approximately unaffected by crosstalk, constituting the ideal measurement scheme. The approximate range of separations, for which SPADE outperforms ideal direct imaging, is determined numerically.

This article is organized as follows. In Section \ref{sec:preliminaries}, we introduce the measurement setting and the necessary tools from estimation theory, as well as the notion of crosstalk. In Section \ref{sec:FI_analysis}, we analyze the impact of crosstalk on the SPADE Fisher information for unbalanced sources. In Section \ref{sec:DI_and_experiment}, we compare the crosstalk-influenced SPADE with direct imaging in experimentally relevant settings. We conclude in Section~\ref{sec:summary}.

\section{Preliminaries} \label{sec:preliminaries}
We begin by introducing the measurement setting and the necessary tools from estimation theory, as well as the notion of crosstalk.

\subsection{Measurement setting}
\begin{figure}[!t]
    \centering
    \includegraphics[width=0.49\textwidth]{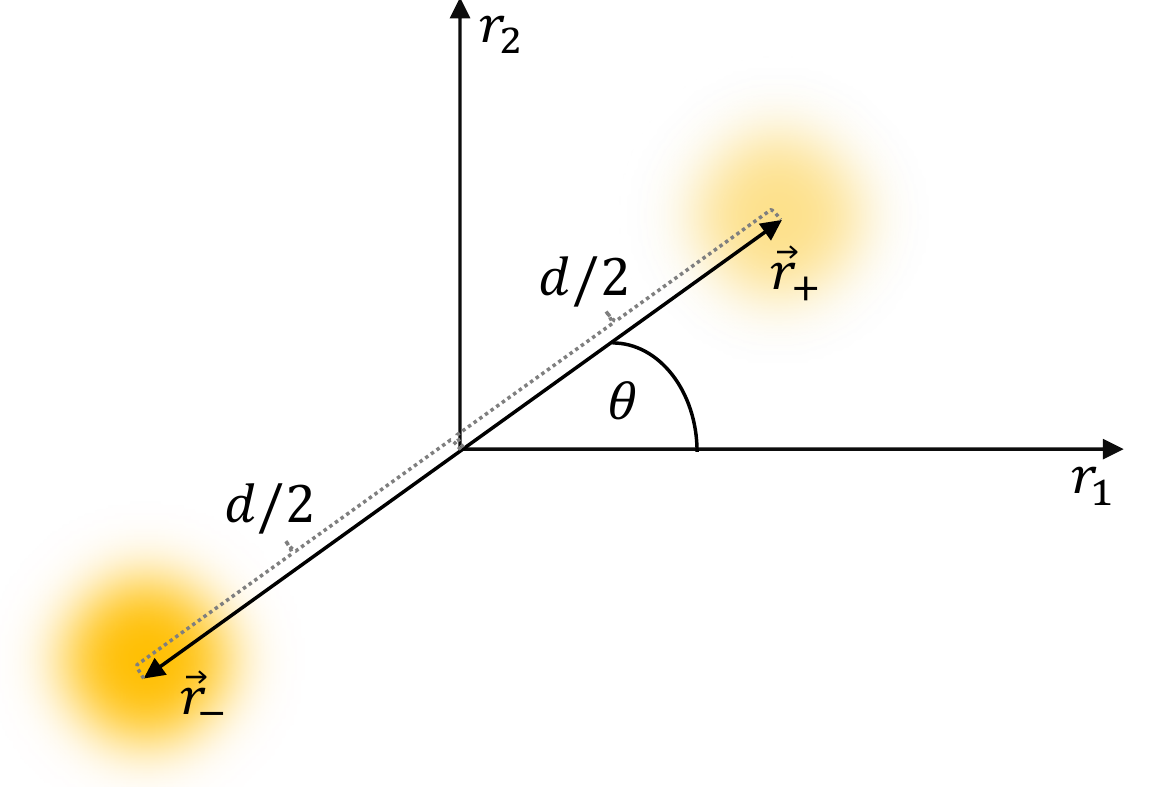}
    \caption{Schematic depiction of the measurement setting. Two incoherent light sources of arbitrary relative brightness separated by distance $d$ lie in the source plane. The origin of the coordinate system in the image plane is centered between them.}
    \label{fig:sources_positions}
\end{figure}

We consider two point-like, incoherent light sources of arbitrary relative brightness following the Poisson distribution, which is the case, e.g. for weak thermal light. We assume that the origin of the coordinate system in the image plane lies between the sources, so that their positions are given by $\vec{r}_\pm=\pm d/2(\cos\theta, \sin\theta)$ for some $d>0$, $\theta\in[0,2\pi)$, as in Figure \ref{fig:sources_positions}.

To quantify the potential imbalance in the brightnesses of the two sources, we define the \emph{relative brightness}
\begin{align} \label{eq:v_definition}
    0 \leqslant \nu \leqslant 1.
\end{align}
Here, $\nu=1/2$ corresponds to equal brightnesses, while $\nu = 1$ ($\nu = 0$) corresponds to the source at $\vec{r}_+$ ($\vec{r}_-$) being the only visible one. For simplicity, further on we assume that the label $\vec{r}_+$ is given to the brighter source, so that $\nu\geqslant 1/2$ (given the opposite scenario, this can be achieved by a simple rotation of the coordinate system in the image plane by $\pi$).\footnote{It is worth adding that the vast majority of our results concerns typical behaviour of the measurement in the presence of crosstalk, which is obtained using either analytical or numerical statistical methods with large sample sizes. Since there is no universally advantageous side of the image plane, in which the brighter source should be placed to achieve better results, the emergent averaged findings are symmetric with respect to the point $\nu=1/2$ anyway.} We remark that in our analysis we treat $\nu$ as a known parameter, i.e. we assume that the relative brightness of the two sources is known prior to the measurement. If this is not the case, our results still provide a meaningful upper bound to SPADE's effectiveness in full multi-parameter estimation, for which the obtained resolution is necessarily smaller \cite{unbalanced_sources_Rehacek_2017}.

The electromagnetic field in the image plane is then conveniently described by two bases centered at the sources: $u_{\pm nm}\coloneqq u_{nm}(\vec{r}-\vec{r}_\pm)$, with $u_{nm}$ being the Hermite-Gauss modes \cite{lasers_1986_siegman}:
\begin{align} \label{eq:HG_modes}
    u_{nm}(\vec{r}) \coloneqq \frac{H_n(\sqrt{2}r_1/w)H_m(\sqrt{2}r_2/w)}{w\sqrt{2^{n+m-1}\pi n! m!}}
        e^{-(r_1^2+r_2^2)/w^2}.
\end{align}
Here, $\vec{r}=(r_1,r_2)$, $H_n(z)\coloneqq (-1)^n e^{z^2} \partial_z^n e^{-z^2}$ are the Hermite polynomials. $w$ is the width of the point spread function of the imaging system, which we assume to be Gaussian. 

Each basis can be used to represent the electric field operator [$\hat{E}^{(-)}(\vec{r})$ is a Hermitian conjugate of $\hat{E}^{(+)}(\vec{r})$]
\begin{align} \label{eq:E}
    \hat{E}(\vec{r}) = \hat{E}^{(+)}(\vec{r}) + \hat{E}^{(-)}(\vec{r})
\end{align}
by expanding its positive-frequency part as
\begin{align} \label{eq:E_+}
    \hat{E}^{(+)}(\vec{r}) 
        = \sum_{nm} u_{+nm}(\vec{r}) \, \hat{b}_{+nm} 
        = \sum_{nm} u_{-nm}(\vec{r}) \, \hat{b}_{-nm},
\end{align}
where $\hat{b}_{\pm nm}$ are the annihilation operators associated with the modes $u_{\pm nm}$. 

The electromagnetic field in the image plane is described by $M$ copies of the quantum state
\begin{align} \label{eq:rho}
\begin{split}
    \hat{\rho}(d) = & \:\nu\hat{\rho}_+ + (1-\nu)\hat{\rho}_-,
\end{split}
\end{align}
with $\hat{\rho}_{\pm}$ being the quantum states of the modes $u_{\pm 00}$, such that \cite{superresolution_Tsang_2016}
\begin{align} \label{eq:epsilon}
\begin{split}
    \Tr \left( \hat{\rho}_{\pm} \hat{b}_{\pm 00}^\dag \hat{b}_{\pm 00} \right) = \epsilon \ll 1.
\end{split}
\end{align}
Consequently, the total number of photons in the electromagnetic field is equal to $N=M\epsilon$.

%It was calculated \cite{crosstalk_original_PRL} that, explicitly,
%\begin{align} \label{eq:f_nm,kl}
%\begin{split}
%    f_{\pm nm,00} &= \frac{1}{\sqrt{n!m!}}\left(\frac{\pm d}{2w}\right)^{n+m} 
%        \cos^n\theta \, \sin^m\theta \, e^{-(d/2w)^2/2}.
%\end{split}
%\end{align}

\subsection{Fisher information and SPADE}

According to estimation theory, the uncertainty $\Delta d$ of estimation of the distance between the sources, based on $N$ measured photons, is determined by the Cram\'{e}r-Rao bound \cite{CR_bound_Cramer_1945,CR_bound_Rao_1945,noon_advances}
\begin{align} \label{eq:Cramer-Rao_bound}
    \Delta d \geqslant 1/\sqrt{N F(d)},
\end{align}
Here, $F$ is the \emph{Fisher information} per photon \cite{Fisher_information_original_1922}, which is a positive quantity dependent on the assumed measurement scheme. The larger the Fisher information, the fewer photons are needed to resolve a given distance.

One of the most widely-used methods of assessing the value of the distance is \emph{direct imaging}, in which the estimation is based on spatially resolved measurements of intensity of light  
\begin{align} \label{eq:I(r)}
\begin{split}
    I(\vec{r}) &\coloneqq 
        M \Tr \left( \hat{\rho}(d) \hat{E}^{(-)}(\vec{r})\hat{E}^{(+)}(\vec{r}) \right)
        %& = N\left(\nu |u_{+00}|^2+(1-\nu) |u_{-00}|^2\right),
\end{split}
\end{align}
in the image plane. While it is a simple and relatively easy-to-implement method, the corresponding Fisher information per photon
\begin{align} \label{eq:FI_DI}
    F_{\textnormal{DI}}(d) = \int_{\mathbb{R}^2} d^2 \vec{r}\frac{1}{p(\vec{r}|d)}
        \left( \frac{\partial}{\partial d} p(\vec{r}|d) \right)^2,
\end{align}
where $p(\vec{r}|d)=I(\vec{r})/N$, is significantly limited due to the presence of diffraction, especially in the \emph{sub-Rayleigh} regime $d/2w<1$.

In principle, the state of light can be measured in any physically implementable basis $v_{nm}$, corresponding to Fisher information per photon equal to
\begin{align} \label{eq:FI_definition}
    F(d) = \sum_{n,m} \frac{1}{p(nm|d)}
        \left( \frac{\partial}{\partial d} p(nm|d) \right)^2,
\end{align}
where $p(nm|d)$ is the conditional probability of detecting a photon in mode $v_{nm}$ when the distance is equal to $d$. An appropriate choice of $v_{nm}$ can result in Fisher information that is larger than the one obtained from direct imaging. By optimizing over all possible measurements, we obtain the \emph{quantum Fisher information}, which corresponds to the smallest uncertainty $\Delta d$ allowed by quantum mechanics \cite{quantum_fisher_information_Braunstein_1994}.
 
In the case at hand, the quantum Fisher information is achieved by spatial mode demultiplexing (SPADE) in the Hermite-Gauss basis centered at the origin, $v_{nm}=u_{nm}$, followed by intensity measurement \cite{superresolution_Tsang_2016}. Remarkably, the corresponding value of the Fisher information per photon is constant:
\begin{align} \label{eq:FI_optimal}
    w^2 F_{\textnormal{HG}}(d) = 1,
\end{align}
both for balanced and unbalanced sources.

Note that in practice, only the first few modes, i.e. those given by $n,m\in\{0,D-1\}$, with $D\geqslant 1$ are measured. The smaller the source separation, the smaller $D$ is required to obtain the ideal value (\ref{eq:FI_optimal}) of the Fisher information. In particular, for distances far below the Rayleigh regime, $d/2w \ll 1$, this ideal value is obtained already by $D=2$ \cite{crosstalk_original_PRL}.

\subsection{Crosstalk} 
In the case of SPADE, as already explained, there is a small but non-vanishing probability that a measured mode is transmitted into an incorrect output. This phenomenon is known as \emph{crosstalk}. Because of crosstalk, the actual measurement basis deviates from the ideal one \cite{crosstalk_original_PRL}:
\begin{align} \label{eq:crosstalk_definition}
    v_{nm}= \sum_{k,l=0}^{D-1} c_{nm,kl}u_{kl},
\end{align}
where we restricted ourselves to $n,m\in\{0,D-1\}$. It is assumed that the diagonal elements of the crosstalk matrix are close to identity and the remaining elements are of a much smaller order, so that, approximately, $v_{nm} \approx u_{nm}$.

Assuming that the contribution from modes given by $n\geqslant D$ or $m\geqslant D$ is negligible, eq. (\ref{eq:crosstalk_definition}) formally defines a change of basis. This implies that the crosstalk matrix is unitary, and as such, can be written as
\begin{align} \label{eq:crosstalk_matrix_unitary_representation}
    c = e^{-i\mu \vec\lambda\cdot\vec{G}},
\end{align}
where $\mu\geqslant 0$, $\vec{\lambda}\in\mathbb{R}^{D^4-1}$ is normalized to one (i.e. $\vec{\lambda}\cdot\vec{\lambda}=1$) and $\vec{G}$ is a vector of all $D^4-1$ generalized Gell-Mann matrices of size $D^2\times D^2$ \cite{Gell-Mann_matrices_1962,Lie_algebras_1999}. For $\mu\ll 1$ the crosstalk matrix is very close to the identity matrix and hence describes small imperfections in the measurement basis:
\begin{align} \label{eq:c_approx}
    c \approx \mathds{1} - i \mu \vec{\lambda} \cdot \vec{G}.
\end{align}

The \emph{crosstalk strength} is defined by the mean off-diagonal matrix element:
\begin{align} \label{eq:crosstalk_strength}
    p_c\coloneqq \frac{1}{D^2(D^2-1)}\sum_{\substack{n,m,k,l=0\\nm\neq kl}}^{D-1} |c_{nm,kl}|^2.
\end{align}
For weak crosstalk matrices (\ref{eq:c_approx}), the crosstalk strength is always proportional to $\mu^2$ and, as we show in Appendix \ref{app:random_crosstalk_matrices}, equals, on average,
\begin{align} \label{eq:p_c(mu)_avg}
    p_c(\mu)\approx \frac{2}{D^4-1}\mu^2.
\end{align}
In practice, we have access to crosstalk strength rather than the abstract parameter $\mu$. For this reason, we base our considerations on $p_c$ wherever possible.

\section{Fisher information in the presence of crosstalk for unbalanced sources}
\label{sec:FI_analysis}
It was shown recently \cite{crosstalk_original_PRL} that while the resolution obtained from SPADE suffers in the presence of crosstalk, it is still higher than that obtained from direct imaging. Here, we investigate how this result generalizes to the case of unbalanced sources.

We start by calculating the corresponding Fisher information. Each detector mode $v_{nm}$ can be associated with its own field operator $\hat{a}_{nm}$. These detector field operators can be expressed as functions of either of the source-centered field operators via
\begin{align} \label{eq:field_operators_detector}
\begin{split}
    \hat{a}_{nm} &= \sum_{kl} f_{\pm nm,kl}(d) \, \hat{b}_{\pm kl},
\end{split}
\end{align}
where
\begin{align} \label{eq:f_nm,kl}
\begin{split}
    f_{\pm nm,kl}(d) &= \int_{\mathbb{R}^2} d^2\vec{r} \, v_{nm}^*(\vec{r}) \, u_{\pm kl}(\vec{r}).
\end{split}
\end{align}
The number of photons in detector mode $v_{nm}$ is then equal to
\begin{align} \label{eq:N_nm}
\begin{split}
    N_{nm} & \coloneqq  M \Tr [\hat{\rho}(d)\hat{a}_{nm}^\dag\hat{a}_{nm}] \\
        & = N \left[\nu|f_{+ nm,00}(d)|^2 + (1-\nu) |f_{- nm,00}(d)|^2\right]
\end{split}
\end{align}
where the bottom line follows from eqs (\ref{eq:rho}, \ref{eq:epsilon}, \ref{eq:field_operators_detector}). 

Therefore, the conditional probability of detecting a photon in the mode $v_{nm}$, under the condition that distance value is $d$, equals
\begin{align} \label{eq:p_SPADE_v}
    p_\nu(nm|d) \coloneqq \frac{N_{nm}}{N} = \nu|f_{+ nm,00}(d)|^2 + (1-\nu)|f_{- nm,00}(d)|^2,
\end{align}
where, in the presence of crosstalk \cite{crosstalk_original_PRL},
\begin{align} \label{eq:gamma_nm} 
    f_{\pm nm,00}(d) = \sum_{k,l=0}^{D-1} c_{nm,kl} \beta_{\pm kl}.
\end{align}
Here,
\begin{align} \label{eq:beta_nm}
    \beta_{\pm kl} = \frac{1}{\sqrt{k!l!}}\left(\pm x\right)^{k+l} 
        \cos^k\theta \, \sin^l\theta \, e^{-x^2/2},
\end{align}
where we introduced the short-hand notation $x\coloneqq d/(2w)$ (note that $x$ is dimensionless).

Thus, the final expression for the SPADE Fisher information per photon for unbalanced sources reads [cf. eq. (\ref{eq:FI_definition})]
\begin{align} \label{eq:FI_definition_unbalanced}
    F_\nu(d) = \sum_{n,m=0}^{D-1} \frac{1}{p_\nu(nm|d)}
        \left( \frac{\partial}{\partial d} p_\nu(nm|d) \right)^2.
\end{align}
Using this formula with eq. (\ref{eq:c_approx}) at the input of eq. (\ref{eq:gamma_nm}), we can compute the Fisher information in the presence of \emph{generic}, i.e. arbitrary unitary, weak crosstalk.

To investigate the behaviour of such Fisher information at small distances, we first observe that in the limit $x \to 0$, the Fisher information is dominated by terms given by $n,m\in\{0,1\}$. In other words, it is enough to consider $D=2$. The corresponding vectors $\vec{\lambda}$, $\vec{G}$ defining the crosstalk matrix via eq. (\ref{eq:crosstalk_matrix_unitary_representation}) are fifteen-dimensional, allowing for explicit analytical treatment. The resulting Fisher information can be then in principle expanded in $x$ to obtain an approximation valid for small separations. 

However, such straightforward expansion will not give us the full picture. To see why, let us consider a specific model of crosstalk given by \emph{uniform} crosstalk:
\begin{align} \label{eq:uniform_crosstalk_single_eq}
    (c_{\textnormal{uni}})_{nm,kl} = \delta_{nk}\delta_{ml} + \left(1-\delta_{nk}\delta_{ml}\right) \sqrt{p_c}.
\end{align}
For such crosstalk matrix, the conditional probability $p_\nu(10|x)$ obtained for small separations $x\ll 1$ and weak crosstalk strength $p_c\ll 1$ reads
\begin{align} \label{eq:p(10|d)}
    p_\nu(10|x) \approx \cos^2\theta \, x^2  + 2\cos\theta(2\nu-1)x\sqrt{p_c} + p_c.
\end{align}
Crucially, the dominating terms on the r.h.s. depend on the relation between the separation and crosstalk strength. For $x\gg \sqrt{p_c}$, the last term is comparatively insignificant, while for $x\ll \sqrt{p_c}$, this is true for the first term. Consequently, the inverse probability entering the Fisher information (\ref{eq:FI_definition_unbalanced}) has two completely different approximate forms:
\begin{align} \label{eq:p(10|d)_expansion}
    \frac{1}{p_\nu(10|x)} \approx \begin{dcases}
        \frac{1}{\cos^2\theta \, x^2}
            \left[1-\frac{2(2\nu-1)\sqrt{p_c}}{\cos\theta\, x}\right] & x\gg \sqrt{p_c}, \\
        \frac{1}{p_c}
            \left[1-\frac{2(2\nu-1)\cos\theta\,x}{\sqrt{p_c}}\right] & x\ll \sqrt{p_c}. \\
    \end{dcases}
\end{align}
To obtain this equation, we simply expanded the l.h.s. in the smallest relevant parameter: $\sqrt{p_c}$ in the top line, and $x$ in the bottom line. For $x\approx \sqrt{p_c}$, neither term in eq. (\ref{eq:p(10|d)}) can be discarded, making it difficult to obtain a simple approximation. 

As seen, if we expand in $x$, we obtain a result that is valid only in the regime $x \ll \sqrt{p_c}$. One can check that this dependence on the ratio between separation and crosstalk strength extends to the other conditional probabilities, and thus the whole Fisher information. Crucially, while above we used the simplified uniform crosstalk model for illustrative purposes, the same qualitative behaviour is observed for any crosstalk. 

Performing the full calculation for the SPADE Fisher information subject to arbitrary crosstalk, we find that
\begin{align} \label{eq:F_limits}
    w^2 F_\nu(x)\approx
    \begin{cases}
        1 & x\gg \sqrt{p_c}, \\   
        q_0 + q_1 x + q_2 x^2 & x\ll \sqrt{p_c},
    \end{cases}
\end{align}
where $q_k$ are functions of $\nu$, $\theta$ and crosstalk. Let us discuss the impact of this result separately in the three ranges $x\gg \sqrt{p_c}$, $x\ll \sqrt{p_c}$ and $x \approx \sqrt{p_c}$.

We start with $x\ll \sqrt{p_c}$. The qualitative effect of crosstalk on the Fisher information in this range (more precise quantitative analysis is provided in the next section) is once again accurately captured by the uniform crosstalk model, for which the coefficients in the bottom line of eq. (\ref{eq:F_limits}) have simple explicit forms:
\begin{align} \label{eq:F_x<epsilon_expansion}
\begin{split}
    q_0 &\approx \left(2\nu-1\right)^2,\\
    q_1 &\approx\frac{2\nu}{\sqrt{p_c}}
        \left(1-\nu\right)\left(2\nu-1\right)\left(\sin^3{\theta} + \cos^3{\theta}\right),\\
    q_2 &\approx\frac{-\nu}{p_c}
        \left(1-\nu\right)\left(4\nu-1\right)\left(4\nu-3\right)
        \left(3 + \cos{4 \theta}\right),
\end{split}
\end{align}
where we restricted ourselves to leading terms in crosstalk strength. In the most radical case of $\nu=1/2$, we can see that $q_0=q_1=0$ and the Fisher information becomes dominated by the quadratic term, making it vanish with decreasing separation as $F_\nu \propto x^2$, a phenomenon originally investigated in \cite{crosstalk_original_PRL}. For other values of $\nu$, the Fisher information $F_\nu$ eventually approaches a positive constant value, which however is always smaller than the ideal one (\ref{eq:FI_optimal}). This shows the influence of crosstalk on the measurement for $x\ll \sqrt{p_c}$ is significant.

In the opposite range, $x\gg \sqrt{p_c}$, we observe a completely different behaviour. There, the Fisher information coincides, to a good approximation, with the optimal value (\ref{eq:FI_optimal}). As such, in this range, crosstalk can be effectively ignored. To see what this means in practice, we performed extensive numerical analysis, finding that, typically, the relation $x\gg \sqrt{p_c}$ corresponds to $x>3\sqrt{p_c}$. In other words, for $x>3\sqrt{p_c}$ the crosstalk-affected SPADE Fisher information is approximately optimal. We remark that for small crosstalk strengths the point $x=3\sqrt{p_c}$ lies deep within the sub-Rayleigh regime $x<1$, meaning that the discussed range is practically relevant. See Appendix \ref{app:x>>pc} for details.

Finally, let us discuss the case of $x\approx\sqrt{p_c}$. Just as in the case of the conditional probability (\ref{eq:p(10|d)_expansion}), in eq. (\ref{eq:F_limits}) we did not provide an approximation for the range $x\approx\sqrt{p_c}$. In this range, an accurate description requires the full expression. To see this, we observe that both lines of eq. (\ref{eq:F_limits}) are extracted from an infinite series in $x$ and $\sqrt{p_c}$ that contains, in particular, terms of the form $(x/\sqrt{p_c})^n$. Clearly, the closer the ratio $x/\sqrt{p_c}$ to one, the fewer terms like these can be discarded, making approximations hard to obtain. Therefore, for $x\approx\sqrt{p_c}$, we study the Fisher information numerically.

\section{Comparison with direct imaging}
\label{sec:DI_and_experiment}
To assess the practical impact of our findings, we compare the crosstalk-affected SPADE with ideal direct imaging in the sub-Rayleigh regime.

We begin our analysis with the asymptotic limit of vanishing distances, $x \to 0$, for which we obtain exact analytical results. From the definition of light intensity (\ref{eq:I(r)}), making use of eqs (\ref{eq:E_+}-\ref{eq:epsilon}), we calculate that
\begin{align}
\begin{split}
    I(\vec{r}|d) = N\left(\nu |u_{+00}|^2+(1-\nu) |u_{-00}|^2\right).
\end{split}
\end{align}
Substituting this into the definition (\ref{eq:FI_DI}) of the Fisher information for direct imaging, expanding the integrand in $x$ and integrating yields
\begin{align} \label{eq:FI_DI_unbalanced}
    w^2 F_{\textnormal{DI}}(x\to 0) \approx (2\nu-1)^2.
\end{align}
This value coincides with the constant term $q_0$ for the SPADE Fisher information (\ref{eq:F_limits}) subject to uniform crosstalk (\ref{eq:F_x<epsilon_expansion}). However, we stress that uniform crosstalk constitutes only a particular model of crosstalk, not necessarily representative of typical crosstalk (indeed, this is exactly what we find below). Estimating the quantitative impact of crosstalk on a typical measurement requires more precise tools. To this end, we instead consider the mean value of the SPADE Fisher information subject to arbitrary crosstalk of a fixed strength. As was the case with uniform crosstalk, in the limit of $x\to 0$, the Fisher information approaches the constant term $q_0$ from the bottom line of eq. (\ref{eq:F_limits}). The goal is to calculate the mean value of $q_0$ averaged over all possible realizations of crosstalk.

Since we are interested only in the limit $x\to 0$, we can once again set $D=2$. We assume that crosstalk matrices are distributed according to eq. (\ref{eq:crosstalk_matrix_unitary_representation}) with $\mu$ related to crosstalk strength via eq. (\ref{eq:p_c(mu)_avg}) and $\vec{\lambda}$ being a 15-dimensional vector distributed according to a normal probability measure. This allows for uniform sampling from the 15-dimensional hypersphere after normalization of $\vec{\lambda}$ \cite{Points_on_sphere}. Substituting the weak crosstalk matrix (\ref{eq:c_approx}) into eq. (\ref{eq:FI_definition_unbalanced}), we find
\begin{multline} \label{eq:q_0_explicit}
    q_{0}\approx\frac{(2\nu-1)^2}{36 n^2}\\
    \bigg[\frac{\sin^2\theta [36 n^2 \alpha_7^2 + 12\mu n\alpha_7\alpha_1(-6 \alpha_{13}+2\sqrt{3}     
        \alpha_{14}+\sqrt{6} \alpha_{15})]}{\alpha_1^2+\alpha_7^2}\\
        +\frac{\cos^2\theta [36 n^2 \alpha_8^2 + 12\sqrt{3}\mu n \alpha_8 \alpha_2(-4 \alpha_{14}+\sqrt{2} \alpha_{15})]}{\alpha_2^2+\alpha_8^2}\bigg].
\end{multline}
where we introduced the parametrization $\alpha_k = n \lambda_k$ with $n=\sqrt{\vec{\alpha}\cdot\vec{\alpha}}$ to account for the normalization of $\vec{\lambda}$ \footnote{We stress that here, e.g., $\alpha_{13}$ means the thirteenth component of the vector $\vec{\alpha}$, i.e. $13$ does not stand for a double index.}. We stress that, being interested in weak crosstalk (\ref{eq:c_approx}), we discard terms of order $\mu^2$ and higher.

We observe that terms linear in $\mu$ are also odd in at least one random variable. Due to the fact that $\alpha_k$ are normally distributed, these terms have vanishing mean values and can be discarded. As a result, we get
\begin{equation}
    q_{0} \to (2\nu-1)^2 \left(\frac{\alpha_7^2\sin^2\theta}{\alpha_1^2+\alpha_7^2}
        +   \frac{\alpha_8^2 \cos^2\theta}{\alpha_2^2+\alpha_8^2}\right).
\end{equation}
We remark that the normalization factor $n$ disappears. The mean value can be now easily obtained by integrating over all random variables $\alpha_k$ with Gaussian weight:
\begin{equation}
   \braket{q_0} = \prod_{k=1}^{15} \int_{-\infty}^\infty \frac{d\alpha_k}{\sqrt{2\pi}}e^{-\alpha_k^2/2} q_0 
    = \frac{1}{2}(2\nu-1)^2.
\end{equation}
The corresponding standard deviation can be calculated in a similar fashion, yielding
\begin{equation}
    \sigma_{q_0}
        \coloneqq\sqrt{\braket{q_{0}^2}-\braket{q_{0}}^2}
        =\frac{1}{4}(2\nu-1)^2\sqrt{\frac{3+\cos 4\theta}{2}}.
\end{equation}
Combining the last two equations, we thus find that the mean Fisher information averaged over generic crosstalk of fixed strength approaches, in the asymptotic limit $x\to 0$,
\begin{align} \label{eq:FI_generic_avg_sd}
    w^2 \braket{F_{\textnormal{gen}}(x\to 0)} \approx 
        \frac{1}{2}(2\nu-1)^2 \left(1 \pm \frac{1}{2}\sqrt{\frac{3+\cos 4\theta}{2}}\right),
\end{align}
where the term after $\pm$ stands for one standard deviation.

\begin{figure}[!t]
    \centering
    \includegraphics[width=0.49\textwidth]{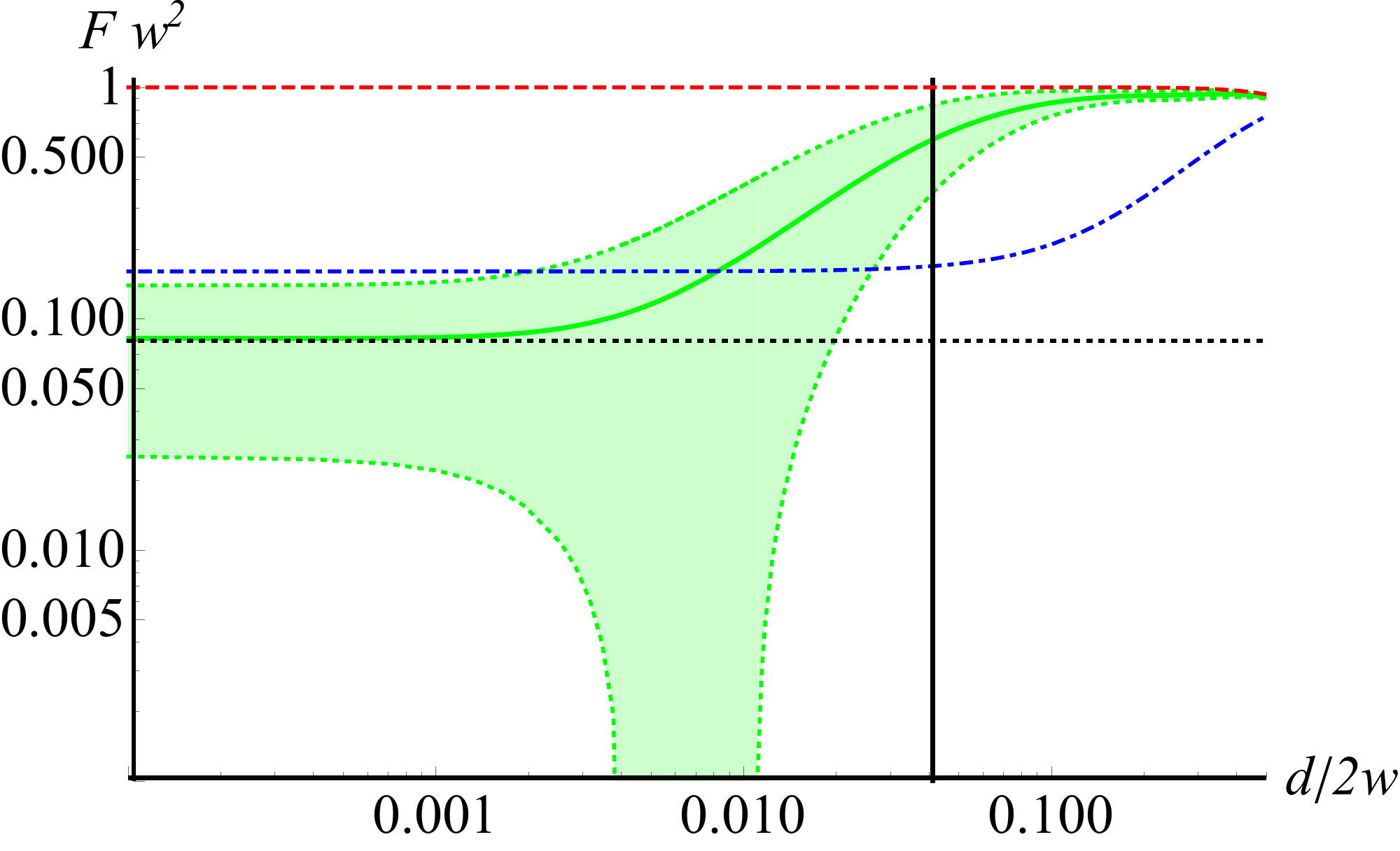}
    \caption{Fisher information for $\nu=0.7$ and $\theta=0$ as a function of measured distance (both quantities conveniently rescaled by the width $w$ of the point spread function). The various lines correspond to direct imaging (blue, dot-dashed), SPADE averaged over 500 random realizations of generic crosstalk of strength $p_c=0.0017$ (chosen after \cite{crosstalk_original_PRL}) (green) with one standard deviation (green bands) and the corresponding analytical asymptotic average (\ref{eq:FI_generic_avg_sd}) (dotted black). In addition, we also plot the optimal measurement value (red, dashed), given by SPADE with no crosstalk, i.e. with the identity substituted for the crosstalk matrix. The SPADE Fisher information was calculated using eq. (\ref{eq:FI_definition_unbalanced}) for $D=3$. As seen, there is a wide range of distances at which crosstalk-influenced SPADE is superior to ideal direct imaging, given by $x>x_c\approx 0.02\sqrt{p_c}\approx 0.008$. In particular, for $x\gg \sqrt{p_c}\approx 0.04$ (denoted by the vertical black line), the crosstalk-averaged Fisher information is nearly at the optimal value (\ref{eq:FI_optimal}) obtained in the absence of crosstalk. It is only in the asymptotic limit of $x\to 0$, where SPADE becomes inferior to noiseless direct imaging.}
    \label{fig:FI_comparison}
\end{figure}

\begin{figure}[!t]
    \centering
    \begin{subfigure}[b]{0.49\textwidth}
         \centering
         \includegraphics[width=\textwidth]{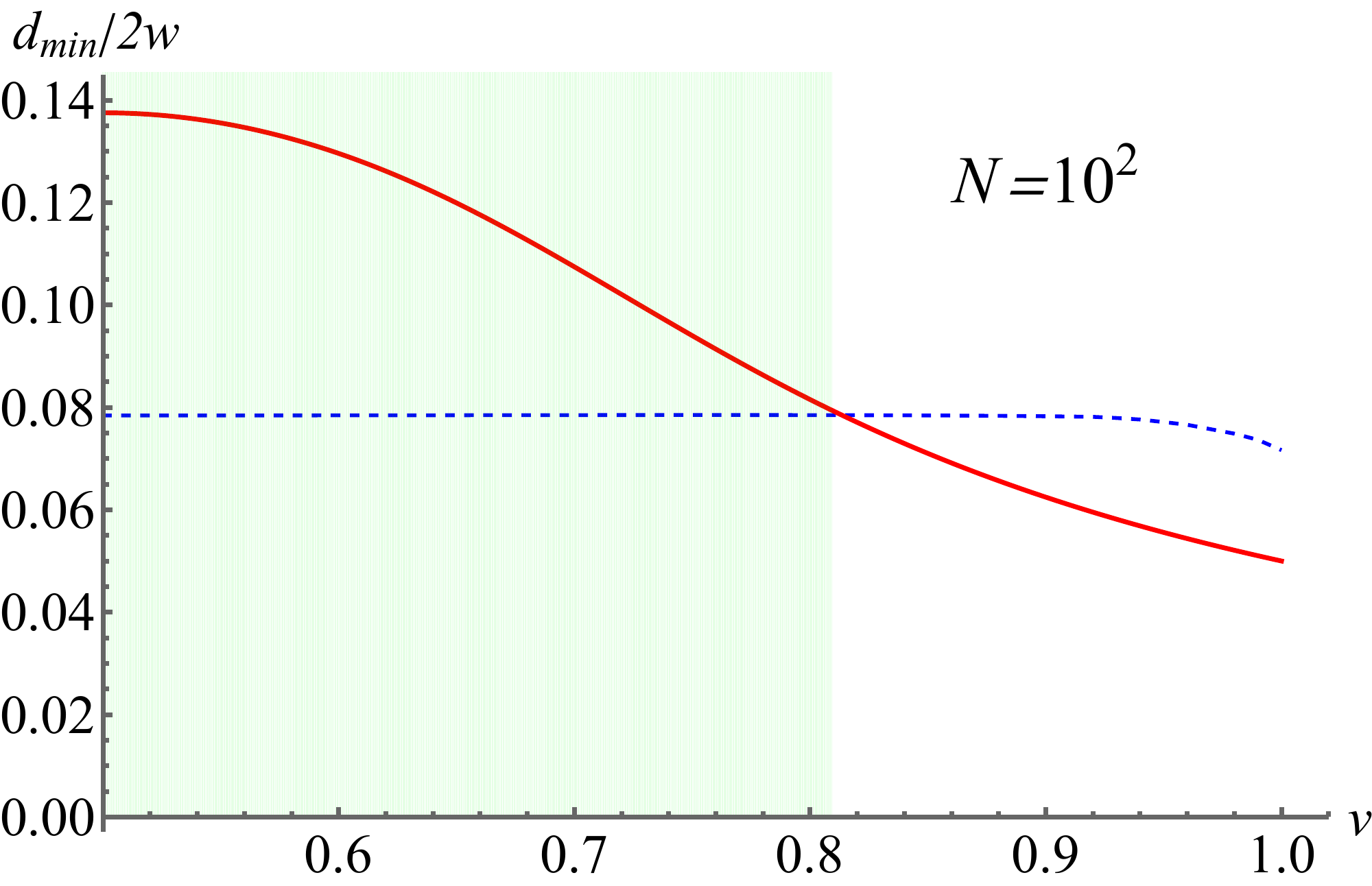}
         \label{fig:experimental_ranges_6}
     \end{subfigure} 
     \hfill\\
     \begin{subfigure}[b]{0.23\textwidth}
         \centering
         \includegraphics[width=\textwidth]{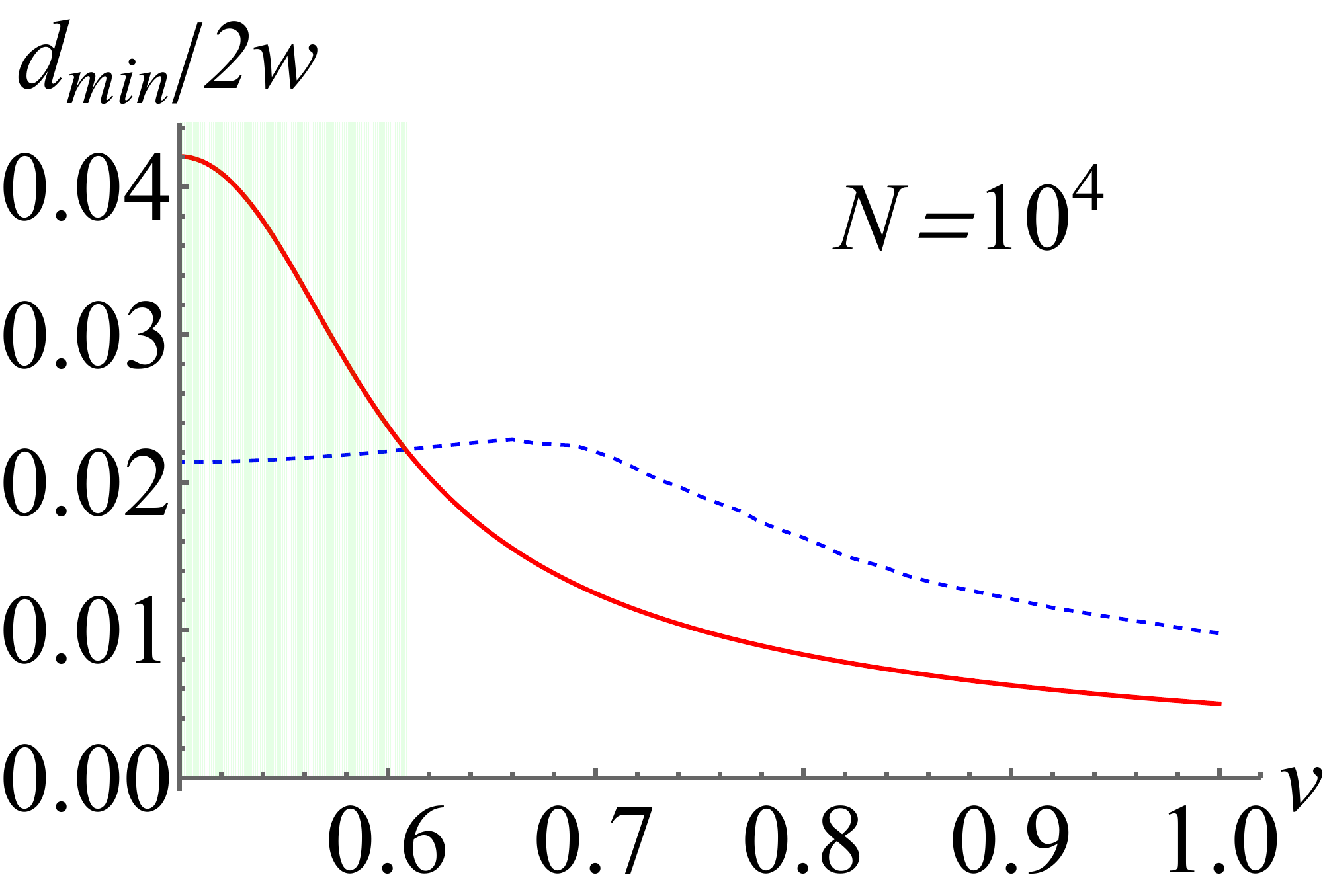}
         \label{fig:experimental_ranges_4}
     \end{subfigure}
     \begin{subfigure}[b]{0.23\textwidth}
         \centering
         \includegraphics[width=\textwidth]{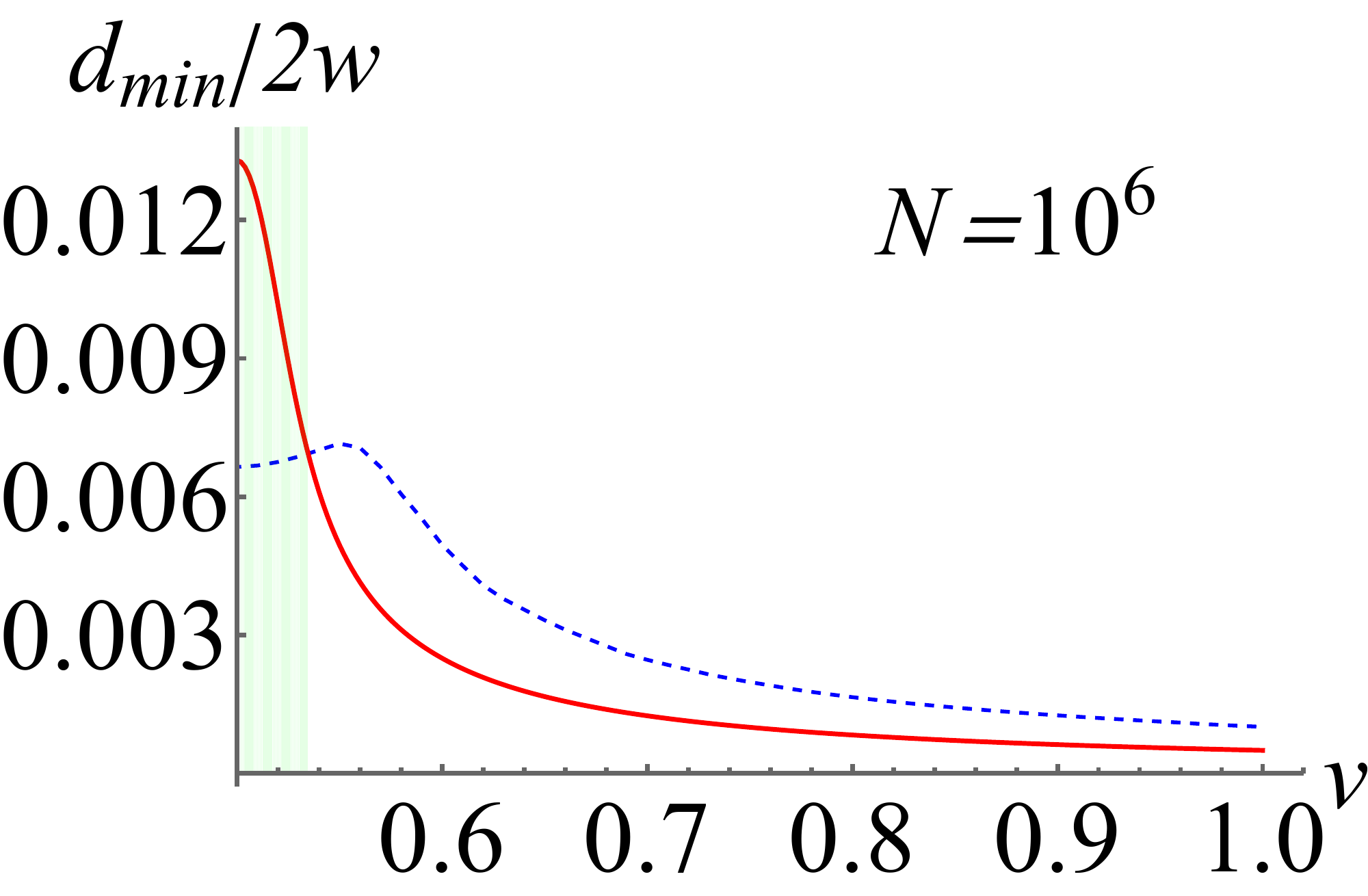}
         \label{fig:experimental_ranges_2}
     \end{subfigure}
        \caption{Minimal resolvable distance for $\theta=0$ and photon numbers $10^2$ (top), $10^4$ (bottom left), $10^6$ (bottom right). The two lines denote: direct imaging (red) and SPADE averaged over 2000 random crosstalk matrices of strength $p_c=0.01$ (blue dashed). Both plots were obtained numerically from definition (\ref{eq:MRD_definition}), where for SPADE we used the Fisher information (\ref{eq:FI_definition_unbalanced}) with $D=3$. As seen, there is always a region (denoted by green colour) beginning at the point $\nu=1/2$, at which SPADE outperforms ideal direct imaging. The smaller the photon number, the larger the width of this region. See main text for additional analysis.}
        \label{fig:experimental_ranges}  
\end{figure}

Comparing this with eq. (\ref{eq:FI_DI_unbalanced}), we find that the average value of crosstalk-influenced SPADE Fisher information is lower than the one obtained from ideal direct imaging even after adding one standard deviation to the former. This result stands in contrast to the case of perfectly balanced sources \cite{crosstalk_original_PRL}, where it was found that SPADE and direct imaging both approach zero for $x\to 0$, with SPADE scaling more favourably. Let us stress, however, that eq. (\ref{eq:FI_DI_unbalanced}) holds for ideal direct imaging only. In reality one should expect direct imaging to perform worse due to finite pixel size and experimental noise, similarly to how SPADE suffers from crosstalk. Thus, in practice, we should expect SPADE to perform better in comparison with direct imaging.

This is indeed what we find when we consider the sub-Rayleigh regime outside of the very limit $x \to 0$. According to our numerical simulations, performed for crosstalk strength in the range $p_c\in[10^{-5},10^{-1}]$ and described in detail in Appendix \ref{app:SPADE_super}, the crosstalk-averaged SPADE Fisher information is larger than its ideal direct imaging counterpart for all separations larger than the \emph{threshold point} $x_c$, with $x_c$ ranging from approximately $x_c \approx 0.05 \sqrt{p_c}$ for $\nu=0.55$ to $x_c \approx 0.2 \sqrt{p_c}$ for $\nu=0.7$. Notably, the threshold point decreases as $\nu$ approaches $1/2$, and in particular equals zero at this point, reproducing known results for balanced sources.

Our findings are illustrated in Figure \ref{fig:FI_comparison}, a comparison between the crosstalk-averaged SPADE Fisher information and the Fisher information for ideal direct imaging. The average over crosstalk was computed numerically by generating vectors $\vec{\lambda}$ according to Gaussian distribution and substiting into eq. (\ref{eq:crosstalk_matrix_unitary_representation}), with $\mu$ related to desired crosstalk strength through eq. (\ref{eq:p_c(mu)_avg}). As seen, in the asymptotic limit $x\to 0$, the crosstalk-influenced SPADE is inferior to ideal direct imaging, confirming our analytical analysis of this region. However, as expected based on our numerical results (described in the previous paragraph), starting from $x_c \approx 0.2 \sqrt{p_c}\approx 0.008$, SPADE becomes superior to direct imaging and in particular, for $x\gg \sqrt{p_c} \approx 0.04$, the crosstalk-averaged Fisher information is nearly at the optimal value (\ref{eq:FI_optimal}) obtained in the absence of crosstalk, as prediced by eq. (\ref{eq:F_limits}).

We can see how the results for Fisher information translate to measurement resolution by considering the \emph{minimal resolvable distance} (MRD) \cite{crosstalk_original_PRL}: the smallest distance between the sources, at which they are distinguishable by the device for a fixed number of measured photons. Mathematically, the MRD is given by the smallest solution $d_{\min}$ to the equation
\begin{align} \label{eq:MRD_definition}
    d_{\min} = 1/\sqrt{N F(d_{\min})}.
\end{align}
This definition follows from the Cram\'{e}r-Rao bound (\ref{eq:Cramer-Rao_bound}) and the fact that one can only resolve distances as small as the estimation uncertainty.

Unfortunately, calculating MRD analytically is typically not feasible. For a single realization of crosstalk, in the asymptotic region $x\approx 0$, the Fisher information may be expanded up to the quadratic term, as in the bottom line of eq. (\ref{eq:F_limits}). It is straightforward to see that the resulting eq. (\ref{eq:MRD_definition}) is equivalent to a quartic polynomial equation in $d_{\min}$, which can be solved using the quartic root formula. However, neither the Fisher information nor the MRD obtained in such a way can be analytically averaged over crosstalk due to the corresponding integrals having non-elementary functional forms. For this reason, we calculate MRD numerically from definition, as before employing the full expression (\ref{eq:FI_definition_unbalanced}) for the SPADE Fisher information for $D=3$.

In Figure \ref{fig:experimental_ranges}, we plot the minimal resolvable distance obtained from crosstalk-affected SPADE and direct imaging as a function of relative brightness. We used photon numbers $N=\{10^2,10^4,10^6\}$ in the orders of magnitude used in contemporary experimental setups \cite{SPLICE_Tham_2017,Rayleigh_curse_Paur_2018,superresolution_Paur_2018} and crosstalk strength $p_c=0.01$, the largest value reported so far \cite{Boucher:20}. From the figure we observe that there is always a non-zero range starting at the point of equal source brightnesses $\nu=1/2$, at which SPADE is superior to ideal direct imaging, despite the fact that the strength of crosstalk is relatively high. Especially at smaller photon numbers SPADE outperforms ideal direct imaging for most values of relative brightness, which suggests it to be an efficient tool for measurements with relatively low number of photons collected. 

From the physical point of view, this behaviour may be seen as a consequence of the fact that, as discussed previously, SPADE becomes superior to direct imaging with growing source separation, where the effects of crosstalk become increasingly negligible. However, large MRD correspond to low photon numbers [this can be seen from eq. (\ref{eq:MRD_definition}) and is also obvious, since fewer measured photons must give us less information]. Hence, for a fixed value of relative brightness, SPADE becomes superior to direct imaging with shrinking photon number, which is exactly what we see in Figure \ref{fig:experimental_ranges}.

\section{Concluding remarks} \label{sec:summary}
We assessed the impact of crosstalk on resolving sub-Rayleigh separations between two unbalanced light sources by spatial demultiplexing. Using statistical methods, we found that the effectiveness of SPADE depends on the relationship between crosstalk strength and the measured separation (normalized with respect to the width of the point spread function). For separations much larger than the square root of crosstalk strength, SPADE still provides the optimal measurement scheme, while for asymptotically vanishing separations its performance is reduced. Numerical simulations performed for realistic values of crosstalk strength and relative brightness show that, while crosstalk-affected SPADE no longer achieves the optimal value of Fisher information, it is still superior to direct imaging for a wide range of experimentally relevant sub-Rayleigh source separations. It is worth noting that the real applicability of SPADE is likely higher due to the fact that we were comparing it to direct imaging with no imperfections.

\begin{acknowledgments}
Project ApresSF is supported by the National Science Centre (no 2019/32/Z/ST2/00017), Poland;  ANR under QuantERA, which has received funding from the European Union's Horizon 2020 research and innovation programme under grant agreement no 731473. M.G. acknowledges funding by MCIN/AEI/10.13039/501100011033 and the European Union “NextGenerationEU” PRTR fund [RYC2021-031094-I]. This work has been founded by the Ministry of Economic Affairs and Digital Transformation of the Spanish Government through the QUANTUM ENIA project call - QUANTUM SPAIN project, by the European Union through the Recovery, Transformation and Resilience Plan - NextGenerationEU within the framework of the Digital Spain 2026 Agenda, and by the CSIC Interdisciplinary Thematic Platform (PTI+) on Quantum Technologies (PTI-QTEP+). This work was carried out during the tenure of an ERCIM ‘Alain Bensoussan’ Fellowship Programme.
\end{acknowledgments}

\bibliography{report}
\addcontentsline{toc}{chapter}{Bibliography}
\bibliographystyle{obib}

\appendix
\section{crosstalk strength and random crosstalk matrices}
\label{app:random_crosstalk_matrices}
\setcounter{equation}{0}
\renewcommand{\theequation}{\ref{app:random_crosstalk_matrices}\arabic{equation}}
\setcounter{figure}{0}
\renewcommand{\thefigure}{\ref{app:random_crosstalk_matrices}\arabic{figure}}
To reliably generate random crosstalk matrices of a given strength, it is necessary to know the relation between crosstalk strength $p_c$ and the number $\mu$ used in the exponential representation (\ref{eq:crosstalk_matrix_unitary_representation}). In this appendix, we derive the relation (\ref{eq:p_c(mu)_avg}) between these parameters and discuss the generation of random crosstalk matrices. 

It follows from eqs (\ref{eq:c_approx}) and (\ref{eq:crosstalk_strength}) that in the lowest order $p_c \propto \mu^2$. In fact, from the construction of the generalized Gell-Mann matrices \cite{Lie_algebras_1999} one can easily calculate that in the lowest order
\begin{align} \label{eq:p_c(mu)}
    p_c(\mu)\approx \frac{2\mu^2}{D^2(D^2-1)}\sum_{k \in S_{nd}}\lambda_k^2,
\end{align}
where the summation is over $k$ corresponding to non-diagonal matrices $G_k$.

To see how $p_c$ typically depends on $\mu$ (when averaged over all possible crosstalk matrices), we assume that each $\lambda_k$ is generated with the same, normal weight. Thus, on average, every $\lambda_k$ must contribute equally to eq. (\ref{eq:p_c(mu)}):
\begin{align}
    \braket{\lambda_k^2} = \frac{1}{D^4-1},
\end{align}
as $\vec{\lambda}$ has $D^4-1$ components. Consequently, since $D^2(D^2-1)$ out of these components correspond to non-diagonal matrices $G_k$, we have
\begin{align}
    \braket{\sum_{k \in S_{nd}}\lambda_k^2} = \frac{D^2(D^2-1)}{D^4-1}.
\end{align}
Taking the average of eq. (\ref{eq:p_c(mu)}) and using the above equation immediately yields eq. (\ref{eq:p_c(mu)_avg}).

\begin{figure}[!t]
    \centering
    \includegraphics[width=0.49\textwidth]{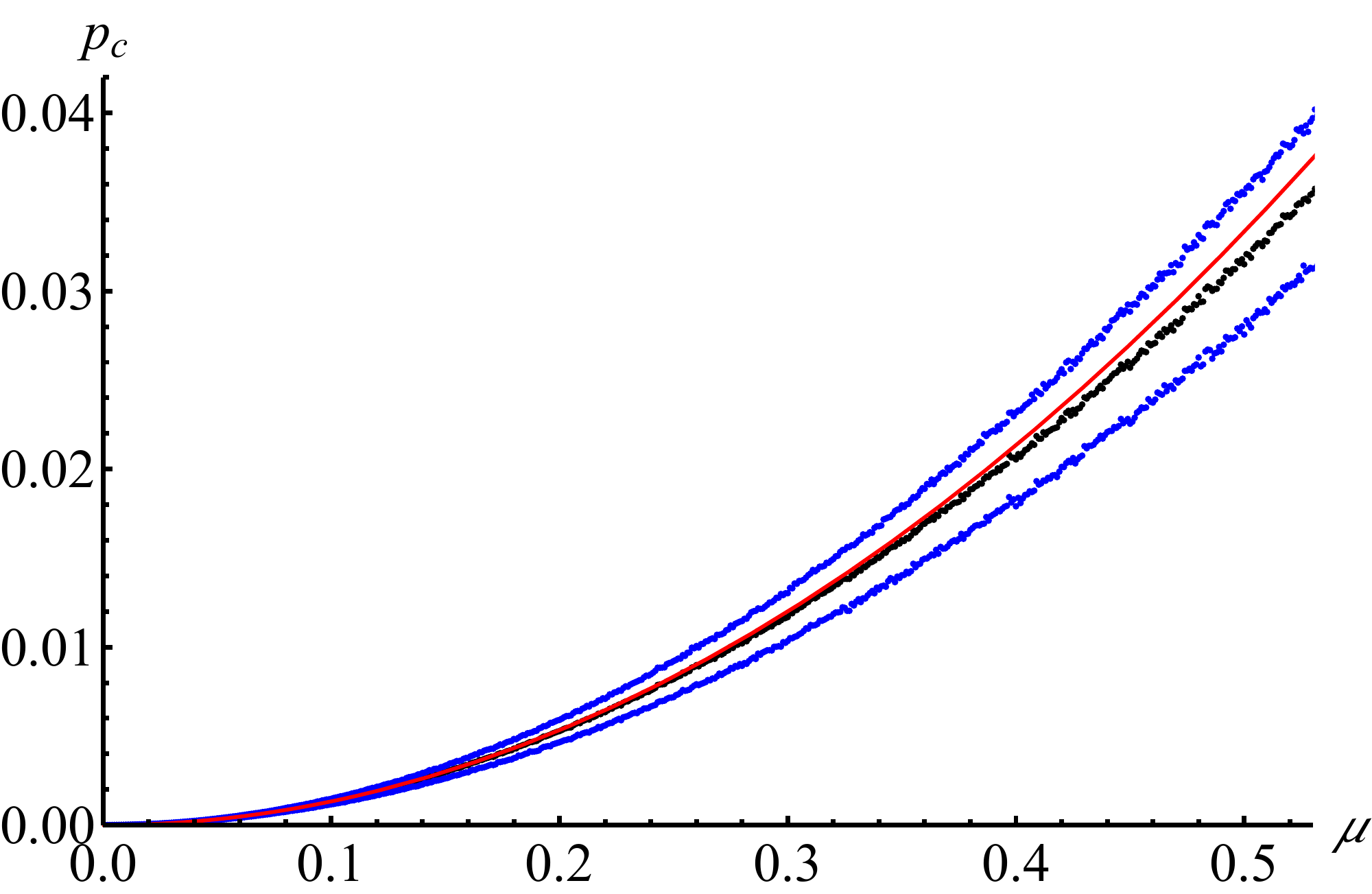}
    \caption{Comparison of the analytical prediction (\ref{eq:p_c(mu)_avg}) for the relationship between crosstalk strength $p_c$ and $\mu$ with numerical data. The analytical prediction is plotted in red, while the mean numerical value is denoted by black circles (in the middle of the plot), with blue circles standing for one standard deviation (top and bottom of the plot). Each data point was obtained by averaging over 500 random generic crosstalk matrices generated for a fixed value of $\mu$.}
    \label{fig:p_c}
\end{figure}

To show that the approximate eq. (\ref{eq:p_c(mu)_avg}) accurately captures the relation between crosstalk strength and $\mu$ for weak crosstalk, we compare the two in Figure \ref{fig:p_c}, finding our approximation and numerical data agree up to one standard deviation.

%For small $\mu$ and thus for weak crosstalks this order should be dominant. Fitting polynomial $a \mu^2 $ to data obtained from random crosstalk matrices for $\mu\in \left[10^{-5},0.9\right]$ we obtain $a\approx2.37\cdot 10^{-2} $ with  coefficient of determination $R^2=0.997$. Thus, quadratic dependence well captures weak crosstalk strengths. Let us observe that $\sqrt{p_c}$ based on our fitted model is one order of magnitude smaller than $\mu$

To generate random generic crosstalk matrices with desired crosstalk strength, we generate vectors $\vec{\lambda}$ according to Gaussian distribution and substite into eq. (\ref{eq:crosstalk_matrix_unitary_representation}) with $\mu$ calculated from eq. (\ref{eq:p_c(mu)_avg}).

\section{Optimal measurement region}
\label{app:x>>pc}
\setcounter{equation}{0}
\renewcommand{\theequation}{\ref{app:x>>pc}\arabic{equation}}
\setcounter{figure}{0}
\renewcommand{\thefigure}{\ref{app:x>>pc}\arabic{figure}}
As discussed in Section \ref{sec:FI_analysis}, the relation $x\gg\sqrt{p_c}$ determines the range of separations at which the measurement apparatus performs with the highest effectiveness in the presence of crosstalk. To make this relation useful for practical predictions, in this appendix, we give it a precise quantitative meaning.

\begin{figure}[!t]
    \centering
    \begin{subfigure}[b]{0.49\textwidth}
         \centering
         \includegraphics[width=\textwidth]{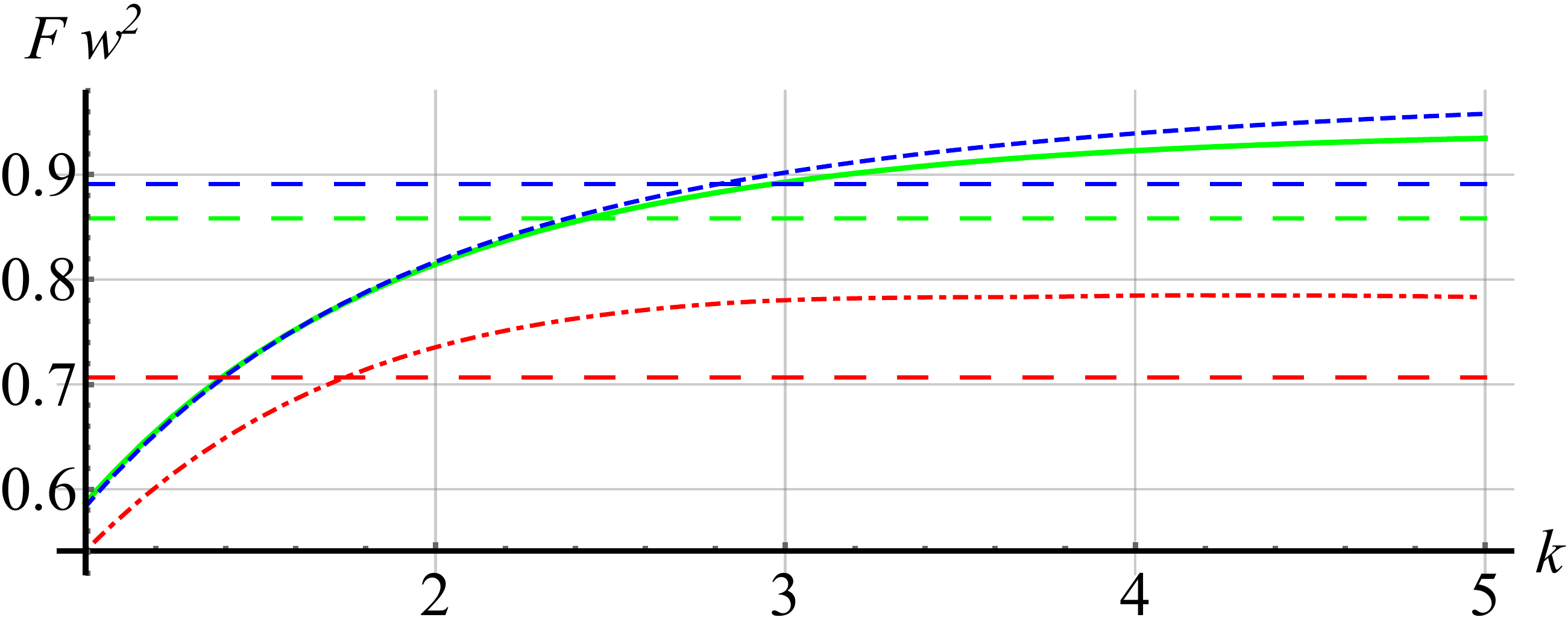}
        \subcaption[]{} \label{fig:Fi_vs_kpc}
     \end{subfigure} 
      \begin{subfigure}[b]{0.49\textwidth}
         \centering
         \includegraphics[width=\textwidth]{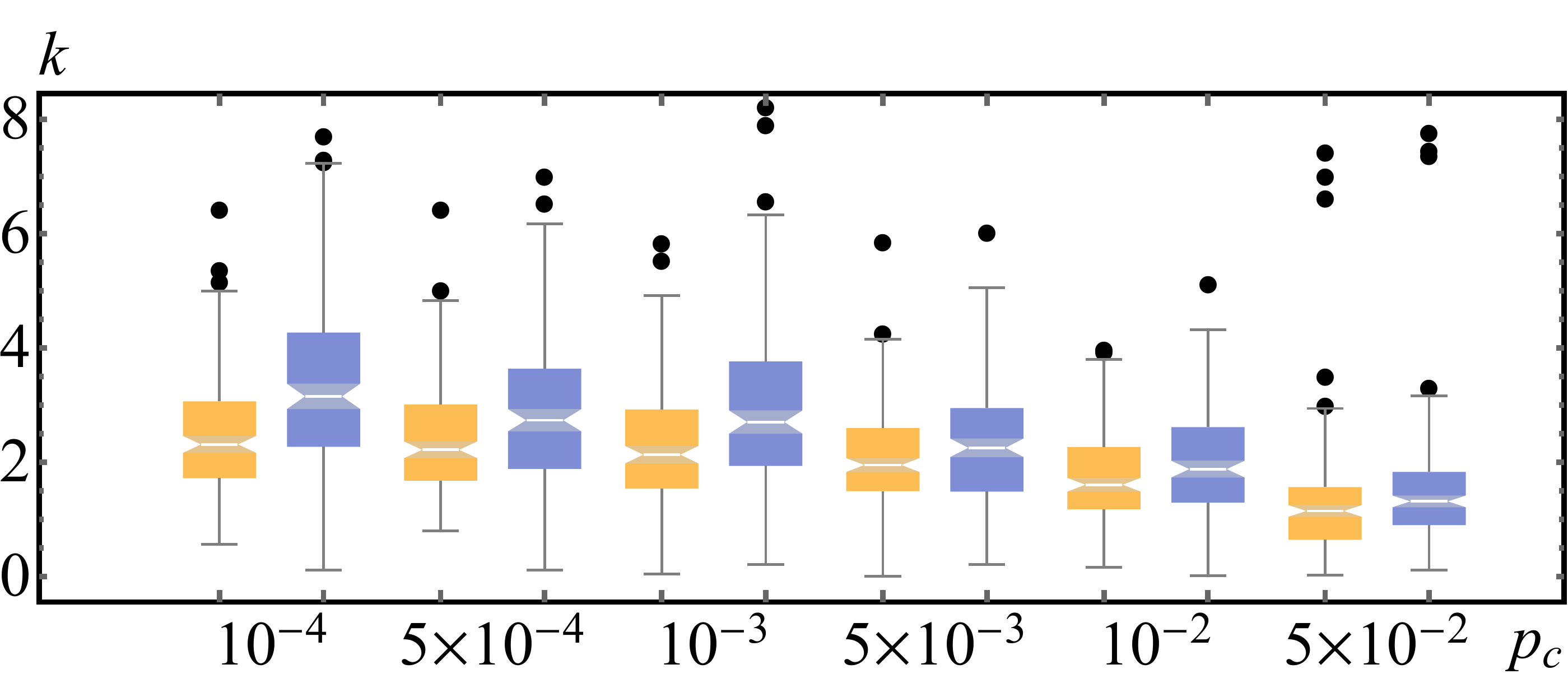}
        \subcaption[]{} \label{fig:median}
     \end{subfigure}
     \begin{subfigure}[b]{0.49\textwidth}
         \centering
         \includegraphics[width=\textwidth]{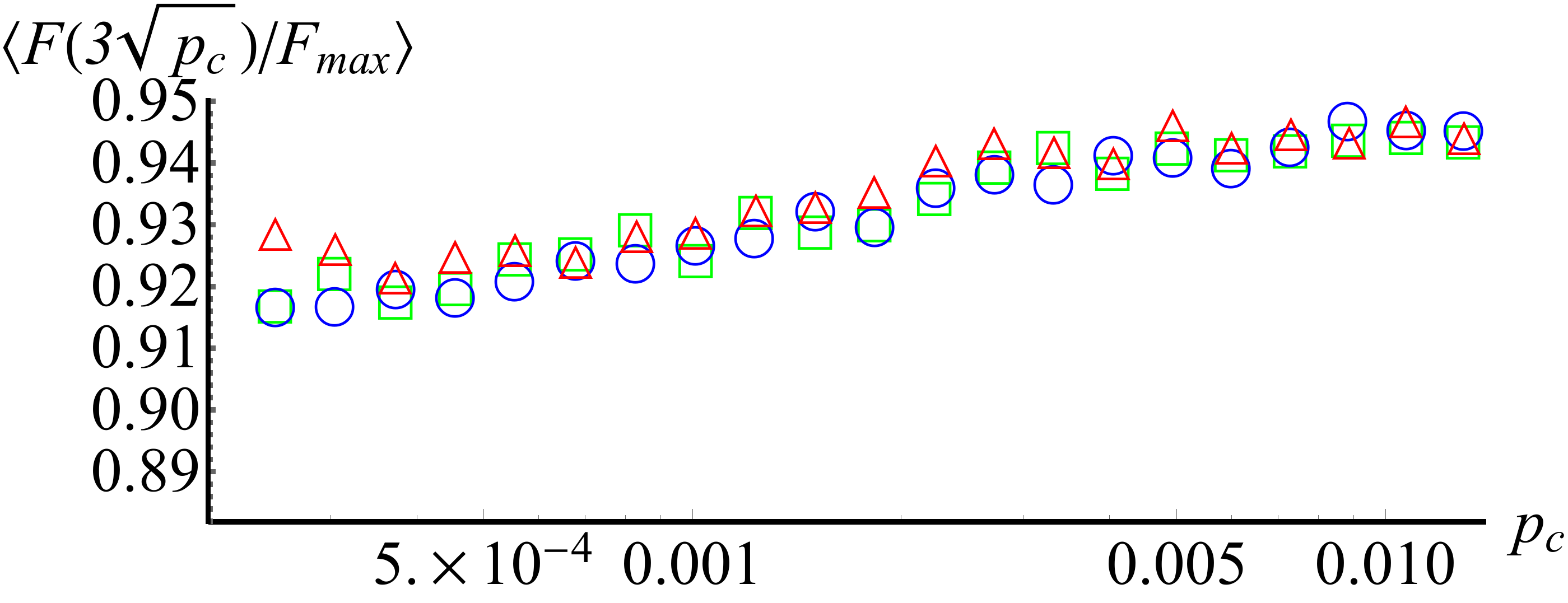}
        \subcaption[]{} \label{fig:percentile_diff}
     \end{subfigure}
        \caption{a) SPADE Fisher information averaged with respect to generic crosstalk as a function of $k=x/\sqrt{p_c}$ for $p_c\in\{10^{-2},10^{-3},10^{-4}\}$ (red dot-dashed, green, blue short-dashed, respectively), $\theta=0$, $D=3$ and $\nu=0.6$. Horizontal dashed lines correspond to $90\%$ of the respective maximal values of Fisher information. b) Box and whisker plot of $k$ required to obtain $90\%$ (yellow brighter boxes) and $95\%$ (blue darker boxes) of maximal Fisher information for various $p_c$ with $\nu$ randomized across samples. Black dots represents outliers. For each box $200$ samples were used. c) Average ratio of Fisher information at the point $x=3\sqrt{p_c}$ and maximal Fisher information as a function of $p_c$ for $\nu \in\{ 0.5,0.6,0.7\}$ (blue circle, green rectangle, red triangle).}
        \label{fig:x>>pc}  
\end{figure}

To this end, we investigate the values of the ratio \begin{align} \label{eq:k}
    k\coloneqq x/\sqrt{p_c}
\end{align}
at which the crosstalk-affected Fisher information is approximately equal to at least $90\%$ of its maximal value. In Figure \ref{fig:Fi_vs_kpc} we present the SPADE Fisher information averaged with respect to generic crostalk as a function of $k$ for $p_c\in\{10^{-2},10^{-3},10^{-4}\}$, $\theta=0$, $D=3$ and $\nu=0.6$. We can see that $90\%$ of the maximal Fisher information is on average approached for the ratio not larger than $k\approx 3$. %However, with decreasing $p_c$ required to obtain such threshold $k$ grows. 

For more in-depth analysis, Figure \ref{fig:median} shows a box plot of $k$ required to obtain $90\%$ and, additionally, $95\%$ of maximal Fisher information for various crosstalk strengths with $\nu$ randomized across samples. We observe that for the case of $90\%$, the median $k$ value is in the approximate range $1.2-2.3$. In the case of $95\%$, the median is in the range $1.3-3.1$. Furthermore, even in the worst cases, the required ratio value does not exceed $k=9$. These results indicate that for typical crosstalk strengths, the relation $x\gg \sqrt{p_c}$ is approximately fulfilled for $x=k\sqrt{p_c}$, with $k$ in the order of unity. Moreover, in the regarded range of crosstalk strengths, for over $75\%$ of considered crosstalk matrices, $k<3$ ($k<4.3$) is enough to obtain $90\%$ ($95\%$) efficiency. 

These results show clearly that, if $90\%$ efficiency is considered sufficient, the optimal measurement region for most crosstalk matrices starts at separations as low as $x=3\sqrt{p_c}$ and is independent of $\nu$. This is additionally supported by Figure \ref{fig:percentile_diff} on which we present the average ratio of Fisher information at the point $x=3 \sqrt{p_c}$ and maximal Fisher information as a function of $p_c$ for $\nu=\{0.5, 0.6, 0.7\}$.

\section{Range of advantage of SPADE over ideal direct imaging}
\label{app:SPADE_super}
\setcounter{equation}{0}
\renewcommand{\theequation}{\ref{app:SPADE_super}\arabic{equation}}
\setcounter{figure}{0}
\renewcommand{\thefigure}{\ref{app:SPADE_super}\arabic{figure}}
In this appendix, we determine the range of separations, at which SPADE is superior to ideal direct imaging. More precisely, we consider the threshold point, or threshold point, $x_c$ for which the Fisher information for both methods is equal.

\begin{figure}[!t]
    \centering
    \begin{subfigure}[b]{0.49\textwidth}
         \centering
         \includegraphics[width=\textwidth]{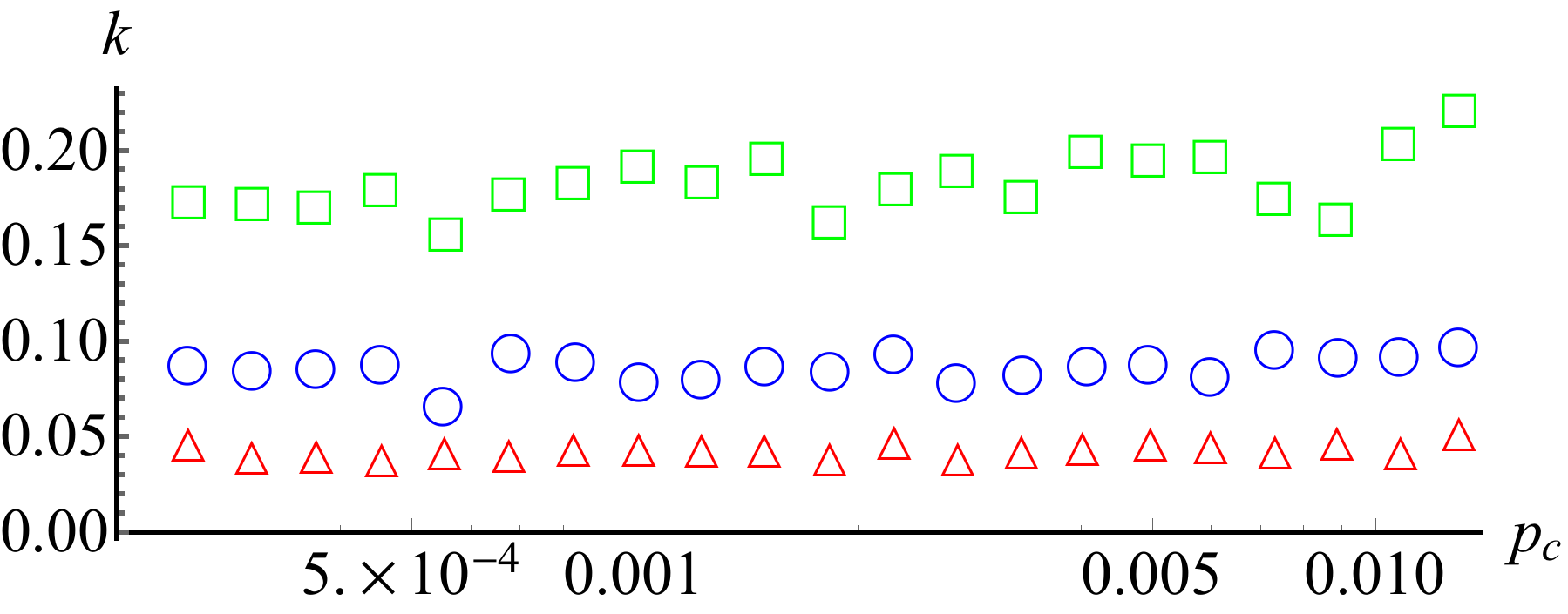}
        \subcaption[]{} \label{fig:trans_1}
     \end{subfigure} 
      \begin{subfigure}[b]{0.49\textwidth}
         \centering
         \includegraphics[width=\textwidth]{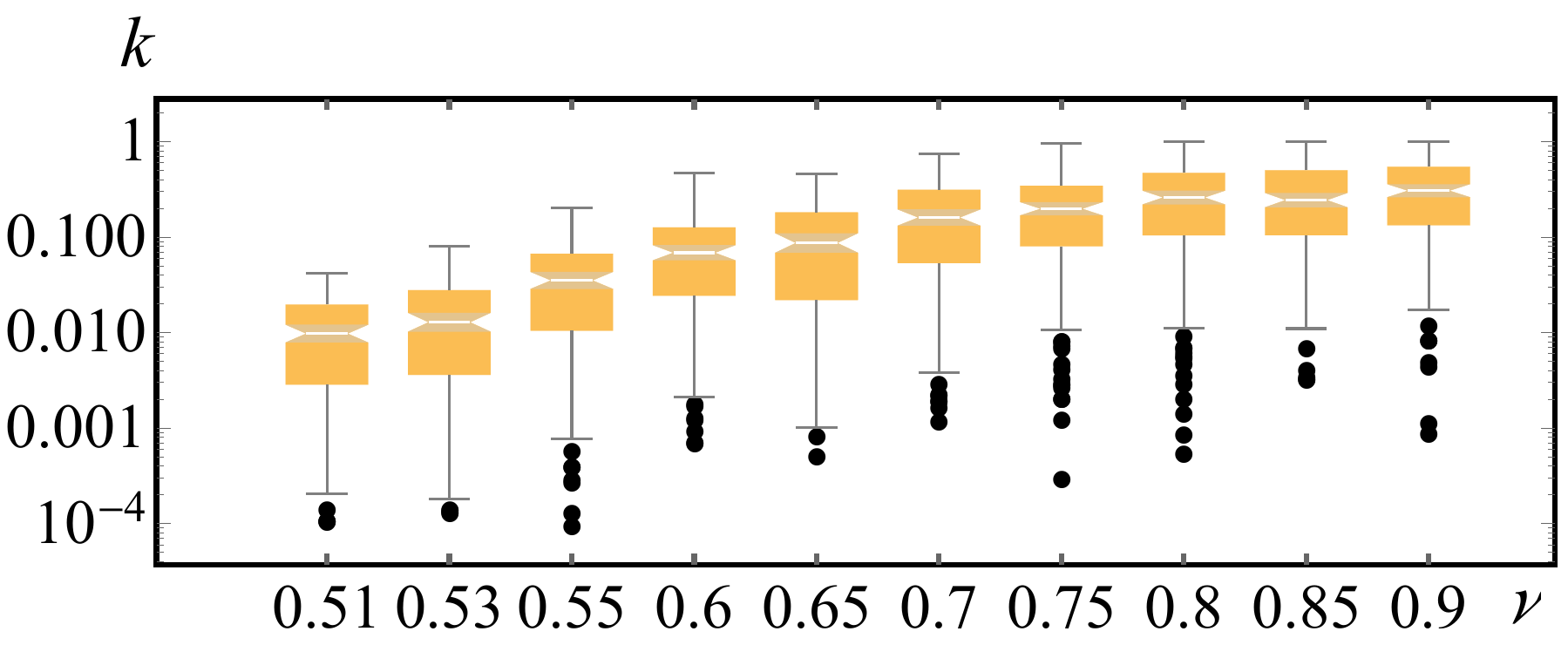}
        \subcaption[]{} \label{fig:trans_2}
     \end{subfigure}
        \caption{a) Mean ratio $k$ corresponding to threshold points averaged over 200 random crosstalk realizations as a function of crosstalk strength $p_c$ for $\nu\in\{0.55,\,0.6,\, 0.7\}$ represented by red triangles, blue circles, green squares respectively. b) Box and whisker plot of $k$ corresponding to threshold points for different $\nu$. For each $\nu$ $200$ crosstalk matrices of random strength were generated. Randomization of $p_c$ was realized by generating a random real $r$ from the interval $[\ln{0.1},\ln{0.8}]$ with uniform distribution and setting $\mu=\exp r$. This roughly corressponds to the range of crosstalk strengths from Figure \ref{fig:trans_1}.}
        \label{fig:trans}  
\end{figure}

%SPADE has an advantage over direct imaging only in some range of $x$. This range is determined by the threshold $x$ for which the Fisher information for both methods is equal. As this point is necessarily in the transition region between the approximations used for Fisher information corresponding to SPADE, one has to use the full expression for Fisher information. Thus, only a numerical analysis of those transition points is easily accessible. 

Once again, we consider the ratio $k$ between separation and crosstalk strength as defined in eq. (\ref{eq:k}). Figure \ref{fig:trans_1} shows the average value of $k$ corresponding to threshold points for 200 crosstalk realizations as a function of the average crosstalk strength for relative brightnesses $\nu\in\{0.55,\,0.6,\, 0.7\}$. We can see that for the three values of $\nu$, the threshold point equals approximately $x_c=\{0.05\sqrt{p_c}, 0.10\sqrt{p_c}, 0.2\sqrt{p_c}\}$, in order of increasing $\nu$. Furthermore, the treshold point has no significant dependence on crosstalk strenght in the considered range. We streess that even for the relatively large relative brightness $\nu=0.7$ the transition point is approximately an order of magnitude smaller than $\sqrt{p_c}$. 

Figure \ref{fig:trans_2} shows a box plot of $k$ corresponding to threshold points for different $\nu$ averaged over random crosstalk of varying strength. From this figure we can see that as $\nu\rightarrow 1/2$, the median of $k$ approaches $0$, which corresponds to SPADE being always better than ideal direct imaging. This is of course expected given the findings for balanced sources, since in such a case there is no threshold point \cite{crosstalk_original_PRL}. In addition, the median decreases faster near $\nu=1/2$. This shows SPADE is especially effective for $\nu$ close to $1/2$ (see also Figure \ref{fig:experimental_ranges}).

%Figure \ref{fig:trans_2} shows a box plot of $k$ corresponding to threshold points for different $\nu$ averaged over random crosstalk of varying strength. From this figure we can see that as $\nu\rightarrow 1/2$ the median of $k\rightarrow 0$, since in such a case there is no threshold point. In addition, the median decreases faster near $\nu=1/2$. This shows that SPADE has its application mostly for $\nu$ near $1/2$ in the cases where $x<\sqrt{p_c}$. For $\nu\geq0.7$ we observe the appearance of cases where $k\approx 1$ however, still most of the results are approximately an order of magnitude smaller than $\sqrt{p_c}$ or smaller. Thus, even in the less optimal scenario for SPADE it has quite a wide range of $x$ where it has an advantage over direct imaging.

\end{document}